\documentclass[letterpaper, 10 pt, conference]{ieeeconf}  
\usepackage{graphicx}
\IEEEoverridecommandlockouts  
\title{\LARGE \bf Respiratory and cardiac monitoring at night using a wrist wearable optical system}

\author{Philippe Renevey$^1$, Ricard Delgado-Gonzalo$^1$, Alia Lemkaddem$^1$, Christophe Verjus$^1$, Selina Combertaldi$^2$,\\ Bj\"orn Rasch$^2$, Brigitte Leeners$^3$, Franziska Dammeier$^4$, and Florian K\"ubler$^4$
    \thanks{$^{1}$Ph.\,Renevey, R.\,Delgado-Gonzalo, A.\,Lemkaddem, and C.\,Vejus are with the Swiss Center for Electronics and Microtechnology (CSEM), Neuch\^atel, Switzerland; e-mail: {\tt\small philippe.renevey@csem.ch}.}%
    \thanks{$^{2}$B.\,Rasch and S.\,Combertaldi are with Department of Psychology of University of Fribourg, Fribourg, Switzerland; e-mail: {\tt\small bjoern.rasch@unifr.ch}.}%
    \thanks{$^{3}$B.\,Leeners is with Universit\"atsSpital Z\"urich, Switzerland; e-mail: {\tt\small brigitte.leeners@usz.ch}.}%
    \thanks{$^{4}$F.\,Dammeier and F.\,K\"ubler are with Ava, Z\"urich, Switzerland; e-mail: {\tt\small florian.kuebler@avawomen.com}.}%
}

\begin{document}

\maketitle
\thispagestyle{empty}
\pagestyle{empty}

\begin{abstract}
Sleep monitoring provides valuable insights into the general health of an individual and helps in the diagnostic of sleep-derived illnesses. Polysomnography, is considered the gold standard for such task. However, it is very unwieldy and therefore not suitable for long-term analysis. Here, we present a non-intrusive wearable system that, by using photoplethysmography, it can estimate beat-to-beat intervals, pulse rate, and breathing rate reliably during the night. The performance of the proposed approach was evaluated empirically in the Department of Psychology at the University of Fribourg. Each participant was wearing two smart-bracelets from Ava as well as a complete polysomnographic setup as reference. The resulting mean absolute errors are 17.4\,ms (MAPE 1.8\%) for the beat-to-beat intervals, 0.13\,beats-per-minute (MAPE 0.20\%) for the pulse rate, and 0.9\,breaths-per-minute (MAPE 6.7\%) for the breath rate.
\end{abstract}

\section{Introduction}
Sleep is a natural mechanism that allows the restoration of cognitive and physical abilities in humans and most mammals. The alteration of normal sleep patterns is an indication of an underlying medical condition or of a degradation of sleep itself, and is therefore a significant indicator of health status~\cite{Steiger2010}. Additionally, sleep is a privileged period during which short or long oscillations of physiological regulation (\textit{e.g.}, circadian rhythms~\cite{Czeisler1999}, menstrual cycles~\cite{Stein2016}, physical training~\cite{Reilly2007}) can be analyzed. At night, exogenous perturbations are limited and thus most of the measured physiological signals are less affected by the environment, allowing analysis of the underlying regulation rhythms.

The gold standard for sleep analysis is the polysomnograph (PSG)~\cite{Brown2012}. The typical setup for polysomnographic measurements involves a plurality of sensors, such as an electrocardiograph (ECG), an electromyograph (EMG), and an electrooculograph (EOG), among others. However, its obtrusiveness makes it unsuitable for long-term sleep studies. Recently it has been shown that using an optical wrist-worn sensor allows to obtain reliable estimates of cardiac activity (\textit{e.g.}, beat-to-beat intervals) with minimal obtrusiveness~\cite{Renevey2013,Parak2015}. Using that information, it is possible to derive several key features such as heart rate, heart rate variability (HRV)~\cite{TFESCNASPE1996}, sleep phases~\cite{Renevey2017}, and respiration rate during sleep~\cite{Aysin2006}.

The current paper presents the performance in terms of accuracy of the wrist-worn device at the following tasks: (1) detecting beat-to-beat intervals, (2) estimating heart rate, and (3) estimating respiration rate. The overall structure of the underlying algorithm is provided as well as the key modules that determine the limits of the system.

\begin {figure}[htbp]
\begin {center}
\includegraphics [width=\columnwidth]{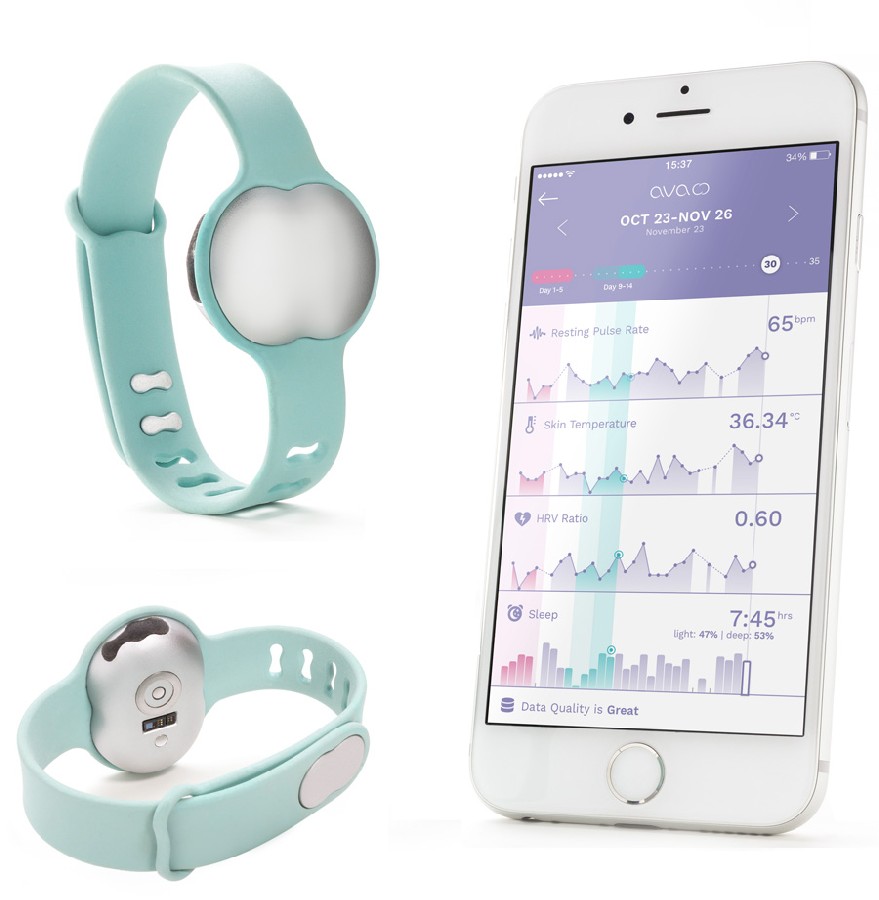}
\caption {\label{fig:ava_proto}Ava smart-bracelet with a multi-wavelength (green and infra-red) photoplethysmographic optical sensor.}
\end {center}
\end {figure}
\section{Materials}

\subsection {Sensors}
The test data was collected using the commercial smart-bracelet developed by Ava\footnote{https://www.avawomen.com} (see Figure~\ref{fig:ava_proto}). This system acquires synchronous data from a three-axis accelerometer and a photoplethysmographic (PPG) sensor on the green and infra-red wavelengths. The sampling frequency for these signals was 25\,Hz. This frequency has been shown to be sufficient to obtain reliable estimates of the beat-to-beat intervals~\cite{Parak2015}. The raw signals were processed within the device to extract beat-to-beat intervals and the average power of the norm of the high-pass filtered acceleration signal. The resulting features are stored on the bracelet and transmitted to a backend server at the end of the night for further analysis and storage.

The reference was composed of a PSG system containing 6 EEG signals (F3, F4, C3, C4, O1, O2), left and right EOG signals, a 1-lead ECG signal, respiration signal and EMG signal. The EEG system that was used was the BrainVision DC Amplifier with GRASS Gold Electrodes. All signals were synchronous and uniformly sampled at a nominal frequency of 200\,Hz. The reference for the beat-to-beat intervals was extracted from ECG and the reference for the breathing rate was recorded using a chest strap.

\subsection {Data acquisition}
The data was collected at the sleep lab of the Department of Psychology of the University of Fribourg. Seven participants, all female, spent four nights each in the lab. In Table~\ref{tb:patients}, we show the subject's statistics. The average interval between consecutive recordings of a participant was about one week. Each participant wore every night two smart-bracelets (one on each wrist), which added up to 56 recordings. As a exclusion criteria during the study, it was stipulated that the wrist sensors should be in contact with the skin at all times, and the reference signals provided by the ECG and the chest strap should be of enough quality to obtain a reliable reference. Thirty-one full-night signals were available after discarding the recordings that did not meet the exclusion criteria. 

\begin{center}
\begin{table}[h]{
\caption{\label{tb:patients}Statistics of the subject's parameters recruited by the Department of Psychology of the University of Fribourg.}}
\hfill{}
\begin{tabular}{c|cccc}
\hline\hline
 & Age (yrs) & Weight (kg) & Height (m) & BMI (kg/m\textsuperscript{2})\\ 
\hline
Mean & 21.14 & 57.21 & 1.69 & 20.03 \\
STD & 1.25 & 7.67 & 0.05 & 1.94\\
\hline\hline
\end{tabular}
\hfill{}
\end{table}
\end{center}

\begin {figure}[htbp]
    \begin {center}
        \includegraphics [width=\columnwidth]{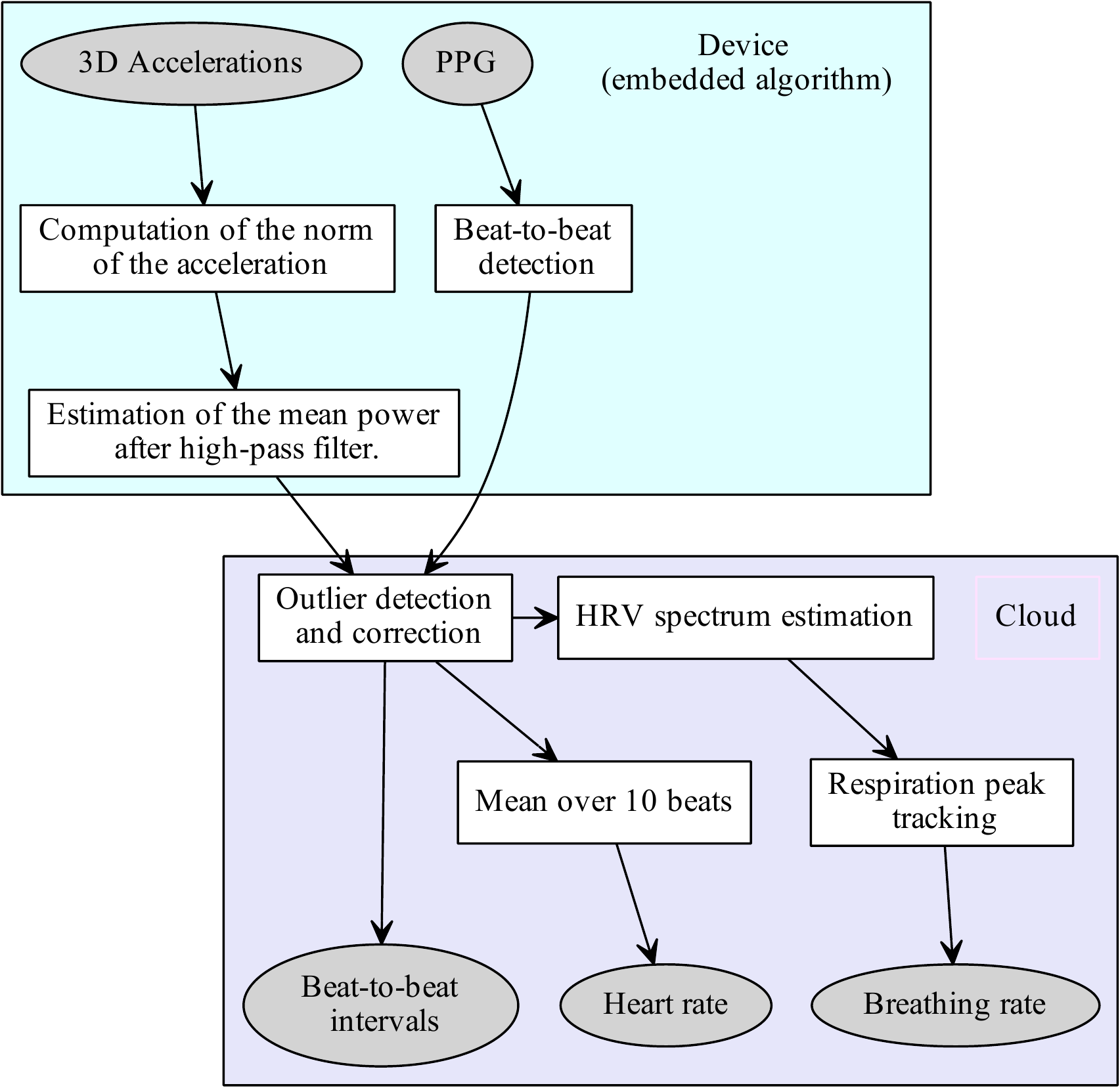}
        \caption {\label{fig:algo}Schematic representation of the proposed system. The algorithm is split between a low-power device and a data cluster in order to balance the overall computational load and memory footprint.}
    \end {center}
\end {figure}

\subsection {Data processing}
The proposed algorithm focuses on the estimation of beat-to-beat intervals, heart rate, and breathing rate, by combining the PPG measurements and the acceleration signals. This algorithm can be divided into two main parts: the first part is performed directly on the device, and the second part is run on a server. The motivation behind this decoupling is to balance the amount of computation that is performed on the bracelet, optimizing for battery life and fast data transfer between bracelet and server. The general outline of the algorithm is shown in Figure~\ref{fig:algo}.

The acceleration signals are used to detect time periods where the optical measurements are corrupted by motion artifacts. In order to find a motion indicator, the norm of the 3D acceleration is first computed, and after the constant gravity component is removed from the norm, the power of the resulting signal is estimated.

The beat-to-beat intervals are calculated by computing the interval of time between consecutive maxima of the first-order derivative of the PPG signal. Physiological constraints, such as the refractory period of the heart, are used to reject false detections that would result in erroneously short intervals. The position of the maxima is determined by fitting a second order polynomial spline, using the values of the maxima sample and the two surrounding samples as interpolation values.

\begin {figure}[htbp]
    \begin {center}
        \includegraphics [width=\columnwidth]{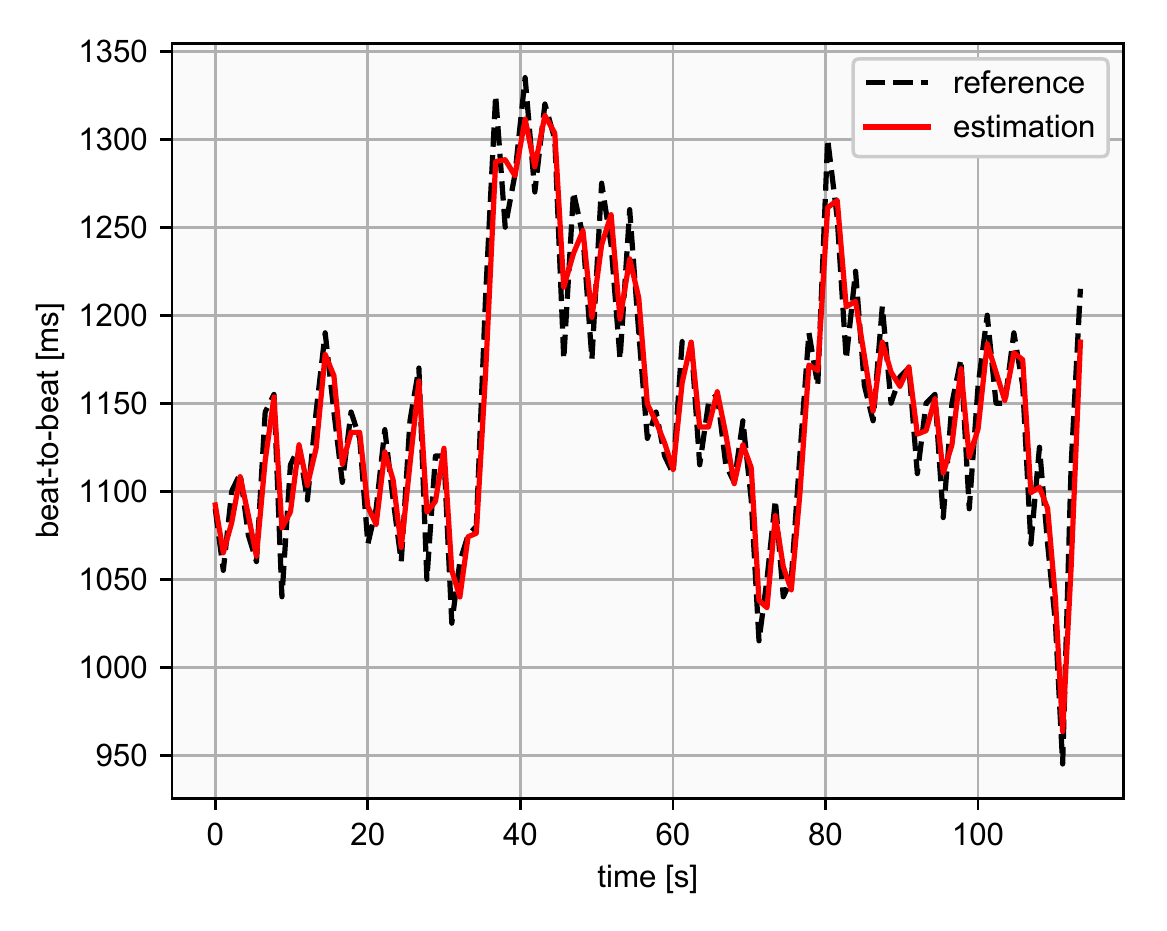}
        \caption {\label{fig:rr_ex}Typical result for the estimation of the beat-to-beat intervals. PSG reference displayed in dashed black and the described approach in solid red.}
\end {center}
\end {figure}

The beat-to-beat intervals that are corrupted by movement or that are not physiologically plausible are corrected using a linear interpolation to replace the corrupted values. 

The heart rate is obtained by averaging the inverse of ten consecutive corrected beat-to-beat intervals. 

\begin {figure}[htbp]
    \begin {center}
        \includegraphics [width=\columnwidth]{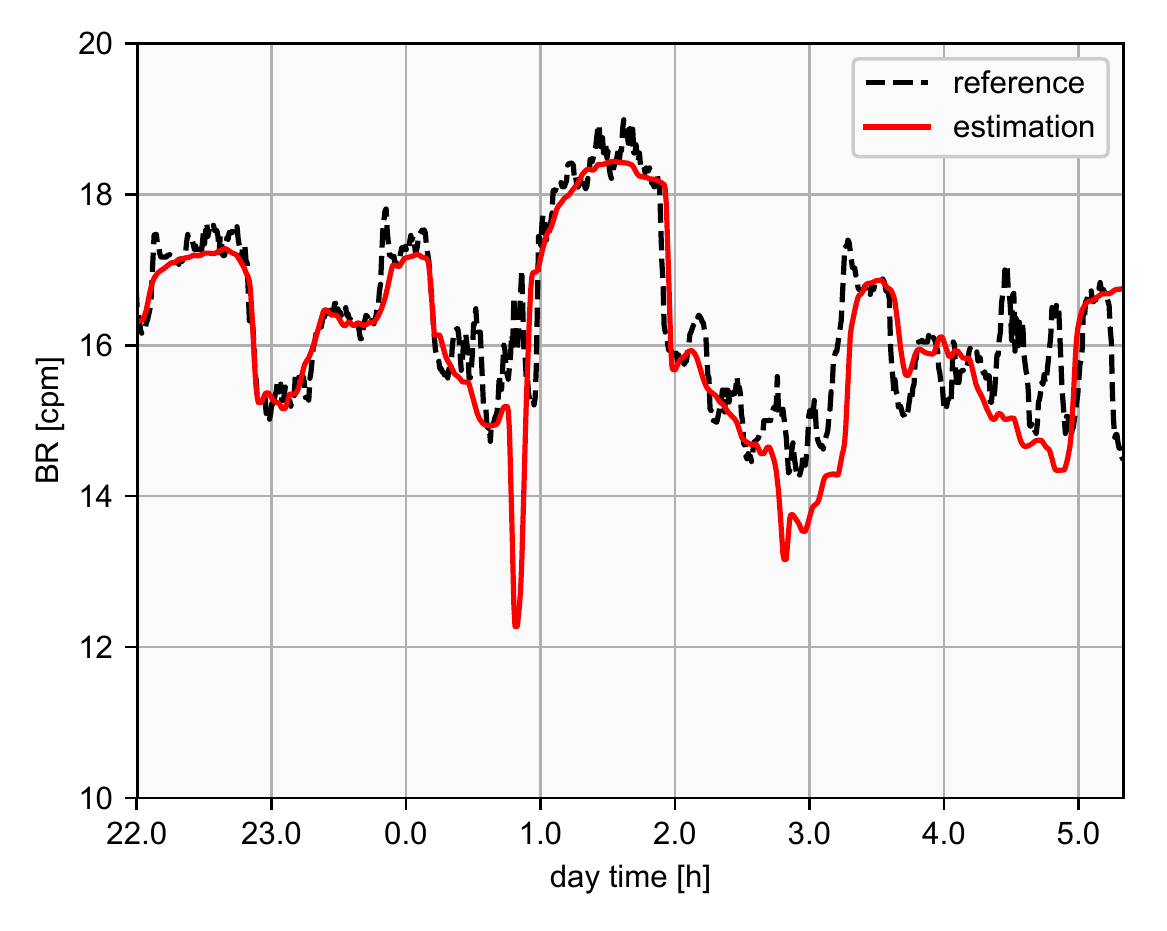}
        \caption {\label{fig:br_ex} Typical result for the estimation of the breathing rate during a full night. In the horizontal axis, the hour of the night is represented. The reference values, from the PSG, are shown in solid black and the estimates of the proposed system are shown in dashed red.}
\end {center}
\end {figure}

The resulting corrected series of beat-to-beat intervals is then uniformly resampled at a frequency of 2\,Hz and band-pass filtered between 0.04 and 0.5\,Hz. This frequency band corresponds to the control of the autonomic nervous system (ANS) over the heart~\cite{TFESCNASPE1996}. The HRV spectrum is obtained by applying an auto-regressive (AR) model~\cite{Haykin2002} of 20$^{th}$ order. The parameters of the AR model are iteratively estimated using a normalized a least mean square algorithm (NLMS)~\cite{Haykin2002}. The resulting HRV spectrum is used to track the peaks in the respiratory frequency band (0.1 to 0.5\,Hz). This band corresponds to respiratory rates between six and thirty breaths per minute, which is the normal range for respiratory rates during the night~\cite{Snyder1964}. The breathing rate is then directly estimated from this peaks frequency. The breathing rate is estimated recursively from the current estimation using the ratio of the power of the respiration peak over the total power in the band as a learning gain. Figure~\ref{fig:br_ex} gives an example of the resulting estimation of the breathing rate (solid red line) compared to the reference value obtained from the PSG's respiration sensor (dashed black line).

\subsection {Performance evaluation}
To determine the accuracy of beat-to-beat intervals, heart rate, and breathing rate estimates from the wrist-worn device, we selected two different performance measures: the mean absolute error (MAE) and the mean absolute percentage error (MAPE).

As the performance evaluation requires a perfect alignment of the test and reference time series, we use a dynamic time warping algorithm (DTW)~\cite{Sakoe1978} to align the respiratory peaks in both datasets. The MAE and MAPE are estimated on the aligned series after the removal of the segments where the reference was corrupted or missing.

The heart rate reference signal is obtained by averaging the inverse of the beat-to-beat aligned intervals over ten beats and the MAE and MAPE is then computed.

To estimate breathing rate accuracy, we interpolated the estimated breathing rate at the times of the reference respiration signal obtained from the PSG setup. The MAE and MAPE are then directly computed from the two series.

\section{Results and discussion}
The results obtained by the proposed approach are presented in Table~\ref{tab:res}. For each evaluation of the performances the minimal (Min), the 25\% quantile (Q$_{25}$), the median, the 75\% quantile (Q$_{75}$), the maximal (Max) and the mean scores are are presented. The different estimations are abbreviated by RR for beat-to-beat interval (in reference to the distance between R peaks within the ECG), HR for heart rate, and BR for breathing rate.

\begin {table}[htbp]
    \caption {\label {tab:res}Performance evaluation for beat-to-beat interval (RR), heart rate (HR), and breathing rate (BR).}
    \hfill{}
    \begin {tabular}{l c c c c c c}
        \hline
        \hline
        & Min & Q$_{25}$ & Median & Q$_{75}$ & Max & Mean\\
        \hline
        RR MAE [ms] & 8.1 & 16.5 & 18.1 & 19.8 & 24. 5 & 17.4\\
        RR MAPE [\%] & 0.7 & 1.5 & 1.8 & 2.0 & 2.4 & 1.8\\
        \hline
        HR MAE [min$^{-1}$] & 0.06 & 0.10 & 0.12 & 0.14 & 0.23 & 0.13\\ 
        HR MAPE [\%] & 0.10 & 0.16 & 0.19 & 0.22 & 0.41 & 0.20\\
        \hline
        BR MAE [min$^{-1}$] & 0.3 & 0.5& 0.8 & 1.2 & 2.5 & 0.9\\ 
        BR MAPE [\%] & 2.1 & 3.8 & 5.5 & 8.7 & 18.2 & 6.7\\
        \hline
        \hline
    \end {tabular}
    \hfill{}
\end {table}

The results obtained for the estimation of the heart rate exhibit a small error (typ. MAE is less than one beat-per-minute) for all the signals. It has to be accounted that the segments where the reference is not reliable have been discarded.

\begin {figure}[htbp]
    \begin {center}
        \includegraphics [width=\columnwidth]{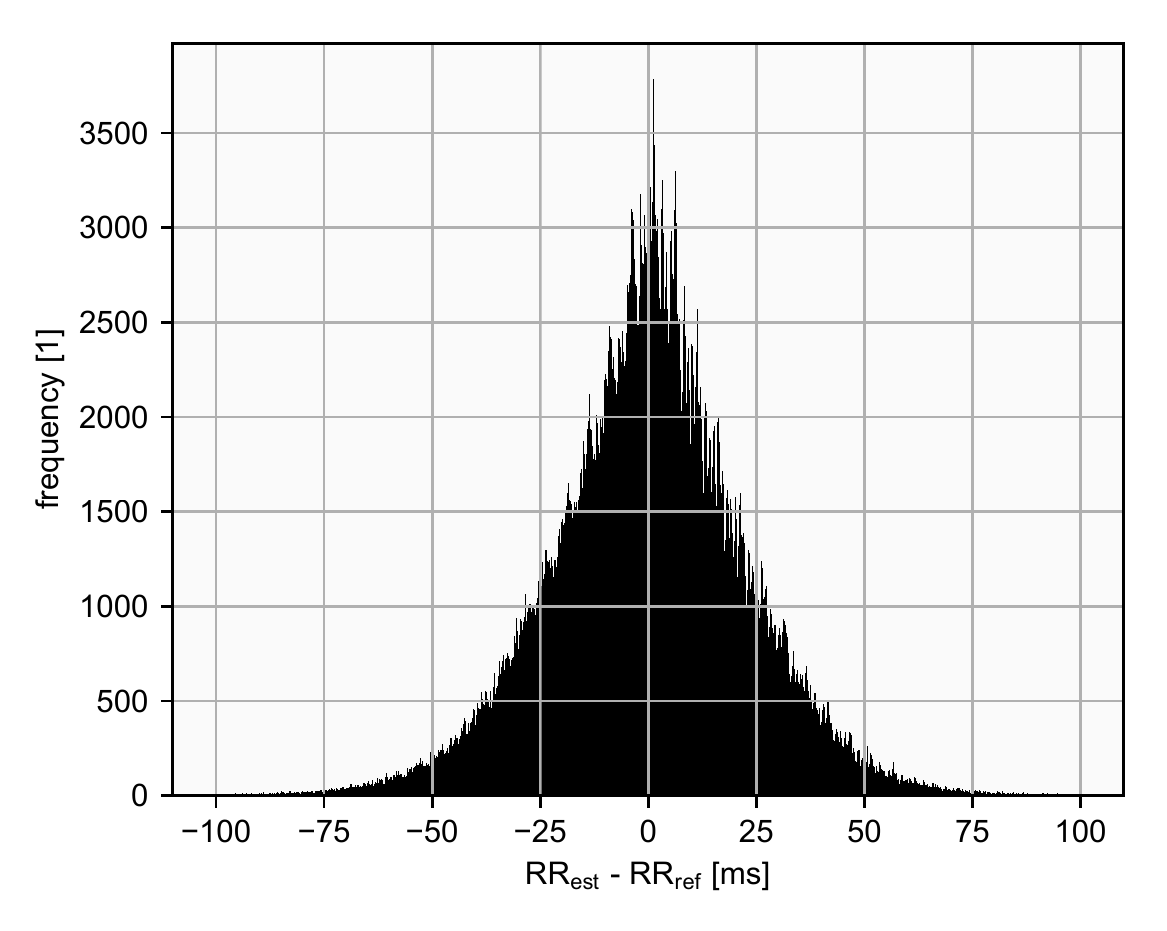}
        \caption {\label{fig:rrh}Histogram of the error of the beat-to-beat detection. The distribution resembles a Laplacian distribution with a mean of 0.08\,ms and a standard deviation of 23.4\,ms)}
\end {center}
\end {figure}

The beat-to-beat intervals exhibit a MAE between 8 to 24\,ms. Such error is acceptable for some applications, such as breathing rate estimates, but can be a limiting factor for other features (\textit{e.g.}, accurate estimation of the power in the HF, LF and VLF spectral bands of HRV spectrum). This error is dependent on different factors. First, the beat-to-beat intervals measured on the wrist are not the exactly identical to the beat-to-beat intervals extracted from the ECG. Indeed the pressure pulse, initiated by the heart contraction, is propagated through the arterial system and undergoes modifications of its shape that affects the accuracy of the arrival time detection. Such distortions are unavoidable and represent a limitation of HRV estimation using a wrist-based sensor. The second factor is the detection algorithm that is embedded in the device. It has to fulfill some constraints about memory and computational complexity, which results in suboptimal detection of the pressure pulses. Improving pulse detection is possible but was out of the scope of this study. Finally, the measured optical signal is strongly sensitive to motion artifacts that can, even for very small movements, affect the accuracy of the pulse wave detection.

The results obtained for heart rate estimates exhibit a small error (typ. MAE is less than one beat-per-minute) for
all the signals. It should be noted that segments with unreliable reference signals, mostly due motion artefacts, have
been discarded. The real value of the MAE for the proposed method is therefore slightly underestimated because the
estimate of the heart rate is also affected by motion artefacts.

The results for the breathing rate show that around 75\% of the signals have a MAE of less than one breath-per-minute. The mean error for the whole dataset is also less than one breath-per-minute. It has to be highlighted that the breathing rate estimates, based on indirect estimation through HRV, require that the participant is in resting condition and that the beat-to-beat series is not strongly corrupted. At the beginning and at the end of the night the person was awake and these two conditions were no longer satisfied, resulting in a much larger error during these time intervals.

\section{Conclusion}
This study has shown that the measurement of beat-to-beat intervals during sleep permit obtaining a reliable estimation of breathing rate. This measurement is made possible because the participant is at rest, allowing a reliable breathing rate detection due to the quasi-absence of motion, and is not applicable to everyday life. For the same reason the heart rate estimation is also accurate. The extraction of beat-to-beat intervals is satisfactory for some applications, but wrist measurements are only an approximation of beat-to-beat intervals obtained from ECG measurements, currently the gold standard for cardiac variability analysis.

It should also be noted that the indirect estimation of the respiration rate from the cardiac variability requires that the ANS driven modulation of the beat-to-beat intervals is normal. For unhealthy or elderly people, whose cardiac function or nervous control of the heart is affected, the proposed approach should be investigated further in a dedicated clinical study.

The algorithms presented in this paper are under development, and in future versions we plan to add supplementary features extracted from the PPG, such as the modulation of amplitude of the observed pulses in the optical signal and the modulation of the baseline. These signal contain relevant information about the breathing rate and the control of the autonomic nervous system, which can then be used to extract relevant information about sleep.

\bibliography{references}

\begin{thebibliography}{10}
\providecommand{\url}[1]{#1}
\csname url@rmstyle\endcsname
\providecommand{\newblock}{\relax}
\providecommand{\bibinfo}[2]{#2}
\providecommand\BIBentrySTDinterwordspacing{\spaceskip=0pt\relax}
\providecommand\BIBentryALTinterwordstretchfactor{4}
\providecommand\BIBentryALTinterwordspacing{\spaceskip=\fontdimen2\font plus
\BIBentryALTinterwordstretchfactor\fontdimen3\font minus
  \fontdimen4\font\relax}
\providecommand\BIBforeignlanguage[2]{{%
\expandafter\ifx\csname l@#1\endcsname\relax
\typeout{** WARNING: IEEEtran.bst: No hyphenation pattern has been}%
\typeout{** loaded for the language `#1'. Using the pattern for}%
\typeout{** the default language instead.}%
\else
\language=\csname l@#1\endcsname
\fi
#2}}

\bibitem{Steiger2010}
A.~Steiger and M.~Kimura, ``Wake and sleep {EEG} provide biomarkers in
  depression,'' \emph{Journal of Psychiatric Research}, vol.~44, no.~4, pp.
  242--252, Mar. 2010.

\bibitem{Czeisler1999}
C.~Czeisler and E.~B. Klerman, ``Circadian and sleep-dependent regulation of
  hormone release in humans,'' \emph{Recent Progress in Hormone Research},
  vol.~54, pp. 97--130, Jan. 1999.

\bibitem{Stein2016}
P.~Stein, L.~Falco, F.~Kuebler, S.~Annaheim, A.~Lemkaddem, R.~Delgado-Gonzalo,
  C.~Verjus, and B.~Leeners, ``Digital womens health based on wearables and big
  data,'' in \emph{Proceedings of the ASRM'16}, 2016.

\bibitem{Reilly2007}
T.~Reilly and B.~Edwards, ``Altered sleep-wake cycles and physical performance
  in athletes,'' \emph{Physiology \& Behavior}, vol.~90, no. 2-3, pp. 274--284,
  Feb. 2007.

\bibitem{Brown2012}
R.~Brown, R.~Basheer, J.~McKenna, R.~Strecker, and R.~McCarley, ``Control of
  sleep and wakefulness,'' \emph{Physiological Reviews}, vol.~92, no.~3, pp.
  1087--1187, July 2012.

\bibitem{Renevey2013}
P.~Renevey, J.~Sol{\`a}, P.~Theurillat, M.~Bertschi, J.~Krauss, D.~Andries, and
  C.~Sartori, ``Validation of a wrist monitor for accurate estimation of {RR}
  intervals during sleep,'' in \emph{Proceedings of the 35th Annual
  International Conference of the {IEEE} Engineering in Medicine and Biology
  Society ({IEEE-EMBS}'13)}, July 3-7, 2013, pp. 5493--5496.

\bibitem{Parak2015}
J.~Parak, A.~Tarniceriu, P.~Renevey, M.~Bertschi, R.~Delgado-Gonzalo, and
  I.~Korhonen, ``Evaluation of the beat-to-beat detection accuracy of {PulseOn}
  wearable optical heart rate monitor,'' in \emph{Proceedings of the 37th
  Annual International Conference of the {IEEE} Engineering in Medicine and
  Biology Society ({IEEE-EMBS}'15)}, August 25-29, 2015, pp. 8099--8102.

\bibitem{TFESCNASPE1996}
{Task Force of The European Society of Cardiology and The North American
  Society of Pacing and Electrophysiology}, ``Heart rate variability standards
  of measurement, physiological interpretation, and clinical use,''
  \emph{Circulation}, no.~93, Mar. 1996.

\bibitem{Renevey2017}
P.~Renevey, R.~Delgado-Gonzalo, A.~Lemkaddem, M.~Proen{\c{c}}a, M.~Lemay,
  J.~Sol{\`a}, A.~Tarniceriu, and M.~Bertschi, ``Optical wrist-worn device for
  sleep monitoring,'' in \emph{Proceedings of the Joint Conference of the
  European Medical and Biological Engineering Conference ({EMBEC}) and the
  Nordic-Baltic Conference on Biomedical Engineering and Medical Physics
  ({NBC}) ({EMBEC\&NBC'17})}, 2017, pp. 615--618.

\bibitem{Aysin2006}
B.~Aysin and E.~Aysin, ``Effect of respiration in heart rate variability (hrv)
  analysis,'' in \emph{Proceedings of the 28th Annual International Conference
  of the IEEE Engineering in Medicine and Biology Society ({IEEE-EMBS}'06)},
  August 30- September 3, 2006, pp. 1176--1179.

\bibitem{Haykin2002}
S.~Haykin, \emph{Adaptive Filter Theory}.\hskip 1em plus 0.5em minus
  0.4em\relax New Jersey, USA: Pearson, 2002, {Fifth edition, 912 p.}

\bibitem{Snyder1964}
F.~Snyder, J.~, Hobson, D.~Morrison, and F.~Goldfrank, ``Changes in
  respiration, heart rate, and systolic blood pressure in human sleep,''
  \emph{Journal of Applied Physiology}, vol.~19, no.~3, pp. 417--422, May 1964.

\bibitem{Sakoe1978}
H.~Sakoe and S.~Chiba, ``Dynamic programming algorithm optimization for spoken
  word recognition,'' \emph{{IEEE} Transactions on Acoustics, Speech, and
  Signal Processing}, vol.~26, no.~1, pp. 43--49, Feb. 1978.

\end{thebibliography}

\end{document}